\documentclass[fontsize=12pt,a4paper,headings=normal,
twoside=false,leqno,parskip=half-,abstract=true]{scrartcl}
\usepackage[english]{babel}
\usepackage[utf8]{inputenc}
\usepackage{hyperref}
\hypersetup{
 pdftitle={global attractors and event horizons},
 pdfauthor={Phillipo Lappicy},
 colorlinks=true,
 linkcolor=blue,
 citecolor=blue,
 filecolor=blue,
 urlcolor=blue}

\usepackage{graphicx}
\usepackage[format=plain,labelfont=bf,font=small]{caption}
\usepackage{xcolor}
\usepackage[arrow, matrix, curve]{xy}
\usepackage{float}

\usepackage{caption}
\usepackage{tabulary}
\usepackage{array}
\newcolumntype{N}[1]{>{\centering\arraybackslash}m{#1}}

\usepackage{amsmath,amsthm}
\usepackage{amssymb} 

\makeatletter
\newcommand{\tpitchfork}{%
  \vbox{
    \baselineskip\z@skip
    \lineskip-.52ex
    \lineskiplimit\maxdimen
    \m@th
    \ialign{##\crcr\hidewidth\smash{$-$}\hidewidth\crcr$\pitchfork$\crcr}
  }%
}
\makeatother
\usepackage{latexsym}

\usepackage[notref,notcite,color,final 
]{showkeys}
\usepackage[pagewise]{lineno}

\definecolor{refkey}{rgb}{1,0,0}
\definecolor{labelkey}{rgb}{1,0,0}

\usepackage{tikz}

\usepackage[textwidth=2cm,textsize=small,backgroundcolor=none]{todonotes}

  \mathchardef\ordinarycolon\mathcode`\:
  \mathcode`\:=\string"8000
  \begingroup \catcode`\:=\active
    \gdef:{\mathrel{\mathop\ordinarycolon}}
  \endgroup

\theoremstyle{plain}
\newtheorem{thm}{Theorem}[section]

\newtheorem{cor}[thm]{Corollary}

\begin{document}

\title{{\LARGE{Space of initial data for 
self-similar Schwarzschild solutions of the Einstein equations}}}

\author{
 \\
{~}\\
Phillipo Lappicy* \\
\vspace{2cm}}

\date{ }
\maketitle
\thispagestyle{empty}

\vfill

$\ast$\\
Instituto de Ciências Matemáticas e de Computação\\
Universidade de S\~ao Paulo\\
Avenida trabalhador são-carlense 400\\
13566-590, São Carlos, SP, Brazil\\


\newpage
\pagestyle{plain}
\pagenumbering{arabic}
\setcounter{page}{1}

\begin{abstract}
The Einstein constraint equations describe the space of initial data for the evolution equations, dictating how space should curve within spacetime. Under certain assumptions, the constraints reduce to a scalar quasilinear parabolic equation on the sphere with various singularities, and nonlinearity being the prescribed scalar curvature of space. We focus on self-similar Schwarzschild solutions. Those describe, for example, the initial data of black holes. We construct the space of initial data for such solutions, and show that the event horizon is related with global attractors of such parabolic equations. Lastly, some properties of those attractors and its solutions are explored.

\ 

\end{abstract}

\section{Main results}

\numberwithin{equation}{section}
\numberwithin{figure}{section}
\numberwithin{table}{section}

The Einstein equations model gravity through spacetime as ten coupled partial differential equations. Six of those evolve space in time, whereas the other four constrain the initial data, or intuitively, dictate how space is curved and embedded in the bigger framework of spacetime. 

We focus on \emph{time-symmetric} spacetime, namely solutions such that the embedding of space in spacetime is trivial and hence its extrinsic curvature vanishes. Hence, those four constraint equations are reduced to only one that indicates how space can bend intrinsically, known as the Einstein's Hamiltonian constraint. See \cite{Rendall07book}. 

Exact solutions of such equation with a prescribed function $T_{00}$ describing its energy density are called \emph{pressureless perfect fluids}, that is, a fluid without pressure, viscosity and heat conduction. See Chapter 4 in \cite{Schutz09book}. They are commonly used in stellar models for idealized distributions of matter, such as stars or black holes. See \cite{Uggla}. 

A simple case among the perfect fluids are the \emph{spherically symmetric} ones. Mathematically, space is described by a three dimensional Riemannian manifold $\mathcal{S}$ with metric $g$. Assume that the space $\mathcal{S}$ can be written in spherical coordinates, that is $\mathcal{S}:=\mathbb{R}_+ \times \mathbb{S}^2$ with $r\in \mathbb{R}_+$ being the radial foliation of two dimensional spheres $(\theta,\phi)\in \mathbb{S}^2$. In the \emph{shear-free} case, the metric splits as 
\begin{equation}\label{METRIC}
    g=u^2 dr^2+r^2 w    
\end{equation}
where $w$ is the standard metric in $\mathbb{S}^2$, and the component $u=u(r,\theta,\phi)$ is the unknown. For a list of known exact spherically symmetric solutions, see Table I and II in \cite{PetarpaVisserWeinfurtner05}.  

Computing the scalar curvature $R(g)$ of $\mathcal{S}$, Bartnik \cite{Bartnik} claimed that $u$ satisfies the following parabolic equation, which fails to be parabolic at $r=0$,
\begin{equation}\label{EEINTRO}
    2ru_r = u^2 \Delta_{\mathbb{S}^2} u + u +\frac{r^2R(g)-2}{2}u^3.
\end{equation}

This parabolic curvature equation was also computed in the appendix of Smith \cite{Smith12}, and is a pure geometric fact of the chosen space $\mathcal{S}$ and metric $g$, and still has no relation to the Einstein equations. This connection is made by prescribing a matter model given by a smooth function $T_{00}$ and relating it with the the scalar curvature $R(g)$ through the Einstein's Hamiltonian constraint equation, 
\begin{equation}\label{EE}
    R(g)=16\pi T_{00}
\end{equation}
as in Rendall \cite{Rendall07book} and Bartnik \cite{Bartnik}.

A simple solution of \eqref{EEINTRO} is given by the Schwarzschild metric, obtained in vacuum, $R(g)\equiv 0$, and when $u$ is independent of the angle variables. This yields a solution $u(r)=\left(1-1/r\right)^{-1/2}$ which blows up at $r_1:=1$. Note that this solution is only valid for $r>r_1$, since for $r\leq r_1$ the metric is not Riemannian. It models the exterior of black holes, where the surface at $r_1$ is known as the \emph{event horizon}, a singularity due to coordinates choice, whereas a \emph{physical singularity} occurs at $r_0:=0$. For a mathematical theory of black holes, see Chandrasekhar \cite{Chandrasekhar83book}.

We seek to understand the interior structure of black holes. In particular, we construct their initial data, as well as relate the interior and the event horizon. The nature of interior of black holes is still debatable. Some possibilities are the constructions of Schwarzschild in equation (35) of \cite{Schwarzschild16}, Synge's formulas (4.2) or (4.6) in \cite{Synge50}, or Florides in equation (2.13) of \cite{Florides74}.

An alternative approach, in order to mimic the exterior Schwarzschild solution to its interior, is to require the same blow-up rate looking from the exterior $(r\to 1^+)$ and interior $(r\to 1^-)$ of the event horizon. This was proposed by Fiedler, Hell and Smith \cite{FiedlerHellSmith}, where they showed that a plethora of angle dependent metrics can occur inside a black hole, with the same horizon. Indeed, plugging the Schwarzschild solution in \eqref{METRIC}, we obtain the metric $g=|1-1/r|^{-1} dr^2+r^2 w$, which is a Riemannian and solves \eqref{EEINTRO} for $r<1$ with a prescribed curvature $R(g)$. For example, if $R(g)=4/r^2$ and the solution is independent of the angle variables, then $u(r)=\left(1/r-1\right)^{-1/2}$ is a solution of \eqref{EEINTRO} for $r<1$. This type of solution models the interior of black holes, in which the metric blows up at the event horizon at $r_1$ with the same rate in the exterior and interior, and has a curvature singularity at $r_0$.

We are interested in \emph{Schwarzschild self-similar interior solutions} of \eqref{EEINTRO}, as
\begin{equation}\label{SSglass}
    u(r,\theta,\phi)=\left(\frac{1}{r}-1\right)^{-\frac{1}{2}}v(r,\theta,\phi)
\end{equation}
for $r<1$. In particular, we construct the space of initial data in order to rigorously study the dynamics of Einstein evolution equations, such as the stability of black holes. The term $(1/r-1)^{-1/2}$ is the interior Schwarzschild blow-up rate of the solution $u$.

Through the self-similar glasses \eqref{SSglass}, $v$ satisfies the following equation for some prescribed scalar curvature $R(g)$, given by \eqref{EE},
\begin{equation}\label{INTRO:IVP0}
    2(1-r)v_r = v^2 \Delta_{\mathbb{S}^2}v-v+\frac{r^2R(g)-2}{2}v^3. 
\end{equation}

Note that the parabolicity of the equation breaks down at the even horizon $r_1:=1$, since there is no radial derivative. Moreover, it is the backwards heat equation for $r>r_1$, which is not well-posed. In order to overcome such problem outside the horizon, Smith \cite{Smith09} uses the coordinates system $u$ satisfying equation \eqref{EEINTRO}.

For $r>r_1$ and certain curvature $R(g)$ were constructed by Smith \cite{Smith09} using the Schwarzschild self-similar exterior solutions of \eqref{EEINTRO} given by
$u=\left(1-1/r\right)^{-1/2}v$. For example, he constructed the metric for certain choices of $R(g)$, and therefore our interior construction can be glued to an exterior solution with such smooth metric. For $r\in(r_0,r_1)$ with $r_0>0$, and curvature $R(g)=(\lambda+2)/r^2$ with $\lambda \in \mathbb{R}_+$, it was shown that there are several non-spherical symmetric solutions in the radial direction bifurcating from the solution $v\equiv 1$, by Fiedler, Hell and Smith \cite{FiedlerHellSmith}.




It is the aim of this paper to study the structure at the event horizon $r_1:=1$ from a dynamical point of view, depending on the interior of a static black hole.

For that, rescale the equation through $r=1-e^{-2t}$, so that the breakdown of parabolicity at $r_1:=1$ is now represented as $t\to\infty$ in 
\begin{equation}\label{INTRO:IVP}
    v_t = v^2 \left[\Delta_{\mathbb{S}^2} v+\frac{r^2R(g)-2}{2}v-\frac{1}{v}\right].
\end{equation}

Note this is a degenerate quasilinear parabolic equation, and hence one can study its initial value problem with initial data at $t=0$, corresponding to $r=0$. 

We are dealing with time independent solutions of the Einstein equations. Even though $t$ is usually called time in parabolic equations, its interpretation here is different: it is a rescaled radial distance from the black hole singularity at $t=0$ such that the event horizon $r_1:=1$ occurs at $t=\infty$.

We recall that horizons occur at spheres in the spatial foliation for some fixed radius, which is a minimal surface such that no other leaf has positive mean curvature, see \cite{Smith09}. Since each leaf $\mathbb{S}^2$ has mean curvature $H=2/(ru)=[2(1-r)^{1/2}]/[r^{3/2}v]$
, horizons occur either at $r_1:=1$ or whenever $v$ is unbounded. 

The main result is now presented: the construction of the space of initial data for such solutions with only one horizon, including the relation of the horizon with global attractors, describing the structure of the metric at $r_1$. 

The equation \eqref{INTRO:IVP} generates a semiflow denoted by $(t,v_0)\mapsto v(t)$ in the phase space $X:=C^{2\alpha+\beta}(\mathbb{S}^2)\cap \{ v>0\}$ where $\alpha,\beta\in (0,1)$ are respectively the H\"older exponent and a fractional power exponent. See \cite{LappicyAxi}. Note that from Lemma 1 in \cite{Smith09}, if $v_0>0$, then $v(t)>0$ for $t>0$ and the space of positive functions is invariant guaranteeing strict parabolicity of \eqref{INTRO:IVP}.

We suppose that the prescribed scalar curvature $R(g)$ is such that the semiflow $v(t)$ is slowly non-dissipative. That is, solutions are global, but they may grow-up and become unbounded as $t\to\infty$. Such solutions would have a different grow-up rate than the Schwarzschild solution. Sufficient conditions for semilinear equations are in \cite{PimentelRocha15}, whereas conditions for quasilinear equations are still not known. Due to such assumption, we disregard blow-up solutions, that is, solutions with other apparent horizons. We consider solutions with a single horizon at $r_1$.

Moreover, slowly non-dissipativity guarantees the existence of an unbounded global attractor $\mathcal{A}$ of \eqref{INTRO:IVP} which attracts all bounded sets as $t\to\infty$. This attractor can be decomposed as the set of equilibria (bounded and unbounded) and heteroclinics between them. See \cite{PimentelRocha15}. 

A particular case is when the dynamical system is dissipative, and solutions stay bounded at $r_1$. Therefore, no grow-up occurs and the solutions have the same grow-up rate as the Schwarzschild solution. Sufficient growth conditions on $R$ for $r\in (0,r_1)$ are
\begin{align}\label{I:DISS}
    R(r,\theta,\phi,v,0) &<\frac{2}{r^2v^2}\nonumber\\
    \left|\frac{r^2v^3}{2}R_{p}\right|\cdot (1+|p|)+\left|\frac{r^2v^3}{2}R-v\right| &<f_1(|v|)+f_2(|v|)|p|^\gamma\\
    \left|\frac{\partial}{\partial_\theta}\left(\frac{r^2v^3}{2}R-v\right)\right|+\left|\frac{\partial}{\partial_\phi}\left(\frac{r^2v^3}{2}R-v\right)\right| &<[f_3(|v|)+f_4(|v|,|p|)](1+|p|)^3\nonumber
\end{align}
where the first condition holds for $|v|$ large enough, uniformly in $(\theta,\phi)$, the second for all $(r,\theta,\phi,v,p)$ for continuous $f_1,f_2$ and $\gamma<2$, the third for $f_3$ nonnegative continuous and monotonically increasing, $f_4$ continuous, monotonically increasing in $|v|$ and tends to $0$ as $|p|\to\infty$ uniformly with bounded $|v|$. $R_p$ Denote both the derivative of $R$ with respect to $v_\theta$ and $v_\phi$. See Chapter 5, section 3 in \cite{Ladyzhenskaya68}. 

Hence, for any bounded initial data $v_0\in X$ at $t=0$ and a scalar curvature $R(g)$ such that $v(t)$ is slowly non-dissipative, there exists a metric $v(t,\theta,\phi)$ in phase space for all $t\in (0,\infty)$. Moreover, if $R$ does not depend on $r$ and $\nabla v$, the solution $v$ will approach an equilibrium in $\mathcal{A}$ as $t\to \infty$, due to the existence of a Lyapunov function
\begin{equation}
    L:=\int_{\mathbb{S}^2}\frac{|\nabla v|^2}{2}-Fd\omega
\end{equation}
where $F$ is the primitive of $-v^{-1}+(r^2R-2)v/2$. This yields that
\begin{equation}
    \frac{dL}{dt}=-\int_{\mathbb{S}^2} \left(\frac{v_t}{v}\right)^2 d\omega
\end{equation}
along trajectories of \eqref{INTRO:IVP}. Note that $v\geq \epsilon >0$ in the phase-space $X$. If $R$ depends on $\nabla v$, there exists a Lyapunov function for axially symmetric solutions, as in \cite{LappicyAxi}. For more general radial foliations, we could possibly obtain a fully nonlinear parabolic equation, instead of \eqref{EEINTRO}. In this case a Lyapunov function for axisymmetric solutions can be available by incorporating the weigth from \cite{LappicyAxi} to \cite{LappicyFiedler18}.

In other words, for any bounded initial data $v_0\in X$ at the singularity $r_0:=0$ of self-similar Schwarzschild solutions, there exists a metric $v(r,\theta,\phi)$ for $r\in (0,r_1)$ such that $v$ converges to an equilibrium of $\mathcal{A}$ as $r\to r_1$. This means that self-similar metrics at the horizon $v(r_1,\theta,\phi)$ are given by equilibria $v_1(\theta,\phi)\in\mathcal{A}$ through $v(r_1,\theta,\phi)=v_1(\theta,\phi)$ and the attractor $\mathcal{A}$ describes the possible metrics at $r_1$. The bounded equilibria $v$ yield models with same blow-up rate as the Schwarzschild solution, whereas the unbounded equilibria yield horizons with different blow-up rate than the Schwarzschild solution.

Then, we use Smith's construction in \cite{Smith09} with such equilibria $v(r_1,\theta,\phi)\in\mathcal{A}$ as initial data at the horizon $r_1$, yielding a metric for $r>r_1$, if one supposes that the scalar curvature $R(g)$ is compactly supported and satisfies 
\begin{equation}\label{DISS2}
    R(g)<\frac{1}{r^2}    
\end{equation}
in $(r_1,\infty)\times \mathbb{S}^2$, and $R(g)=0$ for $[r_1,r_1+\delta)$ for $\delta>0$ small. There is no other horizon for $r>r_1$, 
due to the choice of the standard spherical metric for the foliation.

The above construction shows the following theorem.

\begin{thm}\emph{\textbf{Horizons and Attractors}}\label{I:ATHM1} 

Suppose that space is given by a spherically symmetrc Riemmanian manifold $(\mathcal{S},g)$, that is $\mathcal{S}:=\mathbb{R}_+ \times \mathbb{S}^2$. 

If the scalar curvature $R(g)$ yields a slowly non-dissipative semiflow $v(t)$ of \eqref{INTRO:IVP} satisfying \eqref{DISS2}. Then, for any function $v_0(\theta,\phi)\in X$, there exists a metric for all $r\in\mathbb{R}_+$ given by
\begin{equation}\label{AKI}
    g=\frac{v^2(r,\theta,\phi)}{|\frac{1}{r}-1|}dr^2+r^2\omega
\end{equation}
where $\omega$ is the standard metric on $\mathbb{S}^2$. Moreover, such solutions display only one horizon at $r_1:=1$, which is described by a (possibly unbounded) function $v(r_1,\theta,\phi)$, an equilibrium of the unbounded global attractor $\mathcal{A}$ of \eqref{INTRO:IVP}. 

\end{thm} 

Above, we consider the solutions $v$ which are possibly unbounded at $r_1$, and hence have different grow-up rate comparing to the Schwarzschild solution. If instead of assuming $R$ yields a slowly non-dissipative semiflow, we suppose that it satisfies \eqref{DISS2}, then $v(t)$ is dissipative. Therefore the attractor $\mathcal{A}$ is bounded, and hence are the equilibria within. In this case, solutions have the same grow-up rate as the Schwarzschild solutions.

Recall that the interior region $r\in [0,r_1)$ of the event horizon does not influence the Cauchy development of the exterior $r\in [r_1,\infty)$ of the horizon, see \cite{Wald84}. This is a claim about \emph{time}. The above Theorem is a claim about \emph{space}: the event horizon can not be arbitrary for each fixed time, but it depends on the metric $v$ inside the black hole $r\in [0,r_1)$, in particular at the singularity $r_0=0$. Therefore, the initial data in the horizon can not be freely specified as in Smith \cite{Smith11}, but has additional constraints, namely it has to be within the global attractor of \eqref{INTRO:IVP}.

Therefore, a given shape of the metric at the event horizon impose that only certain possibilities are allowed for the inside of black holes: elements in the basin of attraction of $v(r_1,\theta,\phi)$, i.e., the stable manifold of such equilibrium. Moreover, the evolution dynamics of the Einstein equations of event horizons might constrain even more what is inside a black hole. Even though we can not know for sure what is inside black holes, the above method tells us what can not be inside them.

Bartnik's conjecture, stating that quasi-spherical metrics are typical in the space of smooth metrics, is still an open problem in its full generality. See \cite{BartnikIsenberg}. I also mention, that the quasi-spherical structure is related to the quasi-local mass. Therefore, constructing the phase space for such metrics, might shine a light in the problem of properly defining a suitable quasi-local mass through a dynamical approach. For instance, one can consider the length of the curve $v(r)$ within the stable manifold, between the metrics at $v_0$ and $v(r_1)$.

Also, the metric \eqref{METRIC} is not as general as Bartnik's original \cite{Bartnik}, since we consider shear-free metrics, that is, we do not allow mixed terms $\beta^1 drd\theta$ and $\beta^2 drd\phi$. Metrics with shear satisfy a coupled system, as in \cite{Sharples05}, instead of satisfying a scalar PDE given by \eqref{EEINTRO}, and hence only local existence is proved. It is not known which sufficient conditions yields a dissipative system, as \eqref{DISS2}, and hence global solutions. Also, the system with shear does not necessarily has a Lyapunov function, and the possible metrics at the event horizon are not necessarily equilibria. Therefore, shear-free metrics allow concrete computations of simplified equations  relating global attractors and event horizons. 

The inverse problem is of interest: consider a metric $v(r_1,\theta,\phi)$, which is in $\mathcal{A}$ at the event horizon $r_1$ with prescribed $R(g)$, and find its basin of attraction. Hence, for one given metric at the event horizon, one can find a zoo of possibilities of metrics inside the horizon. Similarly, one can prescribe the metric at the event horizon, and ask which energy density described by the curvature $R$ realizes such equilibria. A similar problem was treated by Rocha and Fiedler \cite{FiedlerRocha99}.


In order to study the existence of other apparent horizons and their interplay, as in \cite{FiedlerHellSmith}, one should drop the slowly non-dissipative assumption and allow blow-up solutions. In this case, the metric $v$ blows-up for $r<r_1$ and other horizons occur inside the event horizon. In such case, a metric between an apparent horizon and an event horizon could be constructed by a heteroclinic within the attractor $\mathcal{A}$.


In other words, we can pin down the space of initial data for the Einstein equations for time-symmetric spherically symmetric self-similar Schwarzschild solutions with one horizon.

\begin{cor}\emph{\textbf{Space of Initial data}}\label{I:SPC} 

Suppose that space is given by a time-symmetric Riemmanian manifold $(\mathcal{S},g)$ with spherical coordinates $\mathcal{S}:=\mathbb{R}_+ \times \mathbb{S}^2$ having shear-free metric $g$ with standard spherical metric $\omega$ in each leaf $\mathbb{S}^2$, and scalar curvature $R(g)$ satisfying \eqref{DISS2}. 

Then, the space of initial data for self-similar Schwarzschild solutions with one horizon at $r_1=1$ is given by $g$ as in \eqref{AKI} such that $v$ lies in the following set
\begin{align}\label{PHASESPACE}
    \mathcal{X}:&=\{v\in C^0([0,\infty),X)\cap C^1((0,\infty),X) \text{ $|$ } v(r_1)\in \mathcal{A}\text{ an equilibrium}\}
\end{align}
where $X:=C^{2\alpha+\beta}(\mathbb{S}^2)\cap \{v>0\}$.
\end{cor} 

The stability of black holes has been widely studied over the past years, as in \cite{DafermosHolzegelRodnianski16}. Usually only linear stability is treated. For the nonlinear stability, the problem is open, and knowing which space the initial conditions belong to, and hence satisfy the constraints, identifies what is rigorously meant by a perturbation of a solution of the Einstein equations, and neighborhoods of solutions in the phase-space \eqref{PHASESPACE}.



Other types of self-similar solutions can be pursued, as Kerr self-similar solutions to model rotating black holes; Reissner$-$Nordstr\"om self-similar solutions to model charged black holes; de Sitter-Schwarzschild self-similar solutions; or a Schwarzschild metric in the exterior of the horizon, and a regular interior in order to model dense stars. Once space of initial data has been constructed for the latter, one can rigorously study the dynamics of stars and its collapse into black holes, as in \cite{Rezzolla04}. 

For the Kerr case of rotating black holes, we consider the following metric
\begin{equation}\label{METRICKERR}
    g=u^2 dr^2+(r^2+a\cos^2(\theta)) w +a\left[1+\frac{r }{r^2+a \cos^2(\theta)} \right]\sin^4(\theta)d\phi^2 
\end{equation}
where $w$ is the standard metric in $\mathbb{S}^2$, $a$ is related to the rotation of the hole, recovering the Schwarzschild case when $a=0$, and the component $u=u(r,\theta,\phi)$ is the unknown. 

We have to find the equation the unknown $u$ satisfies through the scalar curvature of such metric \eqref{METRICKERR}, in case of trace free extrinsic curvature. This will decouple the constraints, so that $u$ satisfies a scalar equation. Note above we only have the spatial metric components, which constrains intrinsic properties of space within the set of initial data. The mixed space and time terms, namely the shift vectors, will come from the extrinsic curvature of the embedding of space in spacetime, that have to solve the momentum constraint equations, constraining the embedding of space within of spacetime. Moreover, we would seek Kerr self-similar solutions of the type $u=[(r^2+a \cos^2(\theta))/(r^2-r+a)]^{-1/2}v$.

On the other hand, note that spatially Reissner$-$Nordstr\"om solutions occur as angle independent solutions of \eqref{EEINTRO} when $R(g)=(4-2q r^{-2})r^{-2}$, where $q\geq 0$ is a constant related to the charge of the black hole, yielding the blow-up solutions $u(r)=\left(r^{-1}-q r^{-2}-1\right)^{-1/2}$ which blows up at $r_\pm:=2^{-1}(1\pm \sqrt{1-4q})$, known as the event horizon at $r_+$, and an interior horizon at $r_-$. Note if $q=0$ we recover the Schwarzschild case. Also, the horizons occur in the same distance, $r_+=r_-=1/2$, when $q=1/4$, and correspond to extremal black holes.

The \emph{Reissner$-$Nordstr\"om self-similar solutions}, namely solutions of \eqref{EEINTRO} of the type
\begin{equation}\label{RNglass}
    u(r,\theta,\phi)=\left(\frac{1}{r}-\frac{q}{r^2}-1\right)^{-\frac{1}{2}}v(r,\theta,\phi)
\end{equation}
will imply that $v$ satisfies the following equation for some prescribed scalar curvature $R(g)$, given by \eqref{EE},
\begin{equation}\label{INTRO:IVP1}
    2[(1-r)-q r^{-1}]v_r = v^2 \Delta_{\mathbb{S}^2}v-[1-q r^{-2}]v+\frac{r^2R(g)-2}{2}v^3. 
\end{equation}

Note that the parabolicity of the equation breaks down at the horizons $r=r_\pm$, since there is no radial derivative. Moreover, strict parabolicity holds for $r(1-r)>q$.

On a similar account, the \emph{Schwarzschild-de Sitter self-similar solutions}, namely solutions of \eqref{EEINTRO} as
\begin{equation}\label{SdSglass}
    u(r,\theta,\phi)=\left(\frac{1}{r}+\frac{\Lambda}{3}r^2-1\right)^{-\frac{1}{2}}v(r,\theta,\phi)
\end{equation}
where $\Lambda$ is the cosmological constant, will imply that $v$ satisfies the following equation for some prescribed scalar curvature $R(g)$, given by \eqref{EE},
\begin{equation}\label{INTRO:IVP2}
    2[(1-r)-\Lambda 3^{-1} r^{3}]v_r = v^2 \Delta_{\mathbb{S}^2}v-[1+\Lambda r^2]v+\frac{r^2R(g)-2}{2}v^3
\end{equation}
where strict parabolicity holds for $3r^{-3}(1-r)>\Lambda$.

Note that in last two equations, modelling Reissner$-$Nordstr\"om and Schwarzschild-de Sitter spaces, even though we can rescale $t(r)$ so that the left-hand side in \eqref{RNglass} and \eqref{SdSglass} have no radial dependence, and become $v_t$, the right hand side will still have radial terms in the linear term in vacuum, for example. Therefore, we still need a deeper understanding of non-autonomous parabolic equations to develop those interior black hole initial data. For example, for such non-autonomous equations we would not have decay to an equilibrium due to the lack of a Lyapunov function. Therefore, it is desirable to understand the dynamics withing global attractors of non-autonomous equations.

Nevertheless, since the Laplacian is still $O(3)$ equivariant, we can use the symmetry breaking methods of Fiedler, Hell and Smith \cite{FiedlerHellSmith}. The problem is that since the equation is nonautonomous, there are no obvious choice of equilibria to bifurcate from.

\color{black}


Also, the Hamiltonian constraint above is one out of four constraints. In the non-time symmetric case, one can rewrite them as a system of equations, as in \cite{Sharples05}, to be studied in the future. 

Due to the no hair theorem, black holes are fully described by their mass, charge and angular momentum. The Schwarzschild self-similarity studied here describes the possible metrics at the event horizon knowing its mass inside, if the black hole is chargeless and has no momentum. The proposals above, namely studying Reissner–Nordstr\"om self-similar solutions, and studying the four constraint equations could describe the full space of initial data for black holes.  

Therefore, we see that the connection between event horizons and global attractors opens many doors yielding new problems to be tackled, with the dynamical perspective of the constraint equations.

In the next section, we explore particular cases of the above theorem, when the global attractor can be explicitely computed, yielding the number of possible equilibria metrics, and certain symmetry within the attractor is known.

\section{Further exploration}

From now on, we expose two corollaries describing properties of the global attractor $\mathcal{A}$ at the event horizon: one describes the possible axisymmetric self-similar Schwarzschild metrics at $r_1$. The other describes some symmetries of certain metric at $r_1$. Both are rigorously proved in \cite{LappicyQuasi}, \cite{LappicyAxi} and \cite{LappicySym}. 

For the first corollary, we are interested into a more detailed study of the structure of the attractor $\mathcal{A}$ that describes the possible metrics at the event horizon $r_1$. For such, we consider axially symmetric solutions and suppose that the metric $v(r,\theta)$ is independent of the angle $\phi\in\mathbb{S}^1$.

Axially symmetric solutions in general relativity have been extensively studied and are also known in the literature as \emph{Ernst solutions} due to \cite{Ernst68}. For a collection of case studies, see \cite{MacCallum11}. Numerical simulation for the dynamics of interaction, pulsation or collapse of axisymmetric stars was done in \cite{Siegel02}. 

Therefore restricting the semiflow to the invariant subspace of axisymmetric solutions $X_{axi}\subseteq {X}$, one obtains a subattractor $\mathcal{A}_{axi}\subseteq \mathcal{A}$ of the flow of
\begin{equation}\label{AXISYMEND}
    v_t = v^2 \left[ v_{\theta\theta}+\frac{v_\theta}{\tan(\theta)} \right]-v+ \frac{r^2 R(g)}{2}v^3
\end{equation}
with Neumann boundary conditions in $\theta\in [0,\pi]$.

In this case, the subattractor $\mathcal{A}_{axi}$ within the axisymmetric subspace $X_{axi}$ can be computed explicitly for any prescribed $R$ satisfying \eqref{DISS2}. This was done for the quasilinear case in \cite{LappicyQuasi} and generalized for the case of singular boundary in \cite{LappicyAxi}. We choose a particular scalar curvature so that the attractor at the event horizon $r_1$ is known.

\begin{cor}\emph{\textbf{A Prescribed Scalar Curvature}} 

If the scalar curvature is given by $R=2r^{-2}[v^{-2}+ \lambda v^3(v-1)(2-v)]$, where $\lambda\in(\lambda_k,\lambda_{k+1})$ and $\lambda_k$ is the $k$-th eigenvalue of the spherical Laplacian with $k\in\mathbb{N}_0$. 
    
Then, the semiflow $v(t)$ is dissipative, the attractor $\mathcal{A}_{axi}$ is compact, and the axisymmetric self-similar Schwarzschild metric at the event horizon $r_1:=1$ is given by one of the $2k+3$ equilibria $v_1,...,v_{2k+3}$ within the Chafee-Infante type attractor in Figure \ref{FIGCOR}, where points denote bounded equilibria and arrows are heteroclinic connections.
\begin{figure}[ht]\centering
\begin{tikzpicture}
\node at (0,0.75) {Schwarzschild self-similar interior solution};
\filldraw [black] (0,0) circle (3pt) node[anchor=south]{$v_{k+2}\equiv 1$};
\filldraw [black] (-1,-1) circle (3pt) node[anchor=east]{$v_{k+1}$};
\filldraw [black] (1,-1) circle (3pt) node[anchor=west]{$v_{k+3}$};

\filldraw [black] (-1,-2.4) circle (0.5pt);
\filldraw [black] (-1,-2.5) circle (0.5pt);
\filldraw [black] (-1,-2.6) circle (0.5pt);
\filldraw [black] (1,-2.4) circle (0.5pt);
\filldraw [black] (1,-2.5) circle (0.5pt);
\filldraw [black] (1,-2.6) circle (0.5pt);

\filldraw [black] (-1,-4) circle (3pt) node[anchor=east]{$v_{2}$};
\filldraw [black] (1,-4) circle (3pt) node[anchor=west]{$v_{2k+2}$};
\filldraw [black] (-1,-5) circle (3pt) node[anchor=east]{$v_{1}\equiv 0$};
\filldraw [black] (1,-5) circle (3pt) node[anchor=west]{$v_{2k+3}\equiv 2$};

\draw[thick,->] (0,0) -- (-0.9,-0.9);
\draw[thick,->] (0,0) -- (0.9,-0.9);

\draw[thick,->] (-1,-1) -- (0.9,-1.9);
\draw[thick,->] (1,-1) -- (-0.9,-1.9);
\draw[thick,->] (-1,-1) -- (-1,-1.87);
\draw[thick,->] (1,-1) -- (1,-1.87);

\draw[thick,->] (-1,-3) -- (0.9,-3.9);
\draw[thick,->] (1,-3) -- (-0.9,-3.9);
\draw[thick,->] (-1,-3) -- (-1,-3.87);
\draw[thick,->] (1,-3) -- (1,-3.87);

\draw[thick,->] (-1,-4) -- (0.9,-4.9);
\draw[thick,->] (1,-4) -- (-0.9,-4.9);
\draw[thick,->] (-1,-4) -- (-1,-4.87);
\draw[thick,->] (1,-4) -- (1,-4.87);

\end{tikzpicture}
\caption{Global attractor $\mathcal{A}$ of Chafee-Infante type} \label{FIGCOR}
\end{figure}
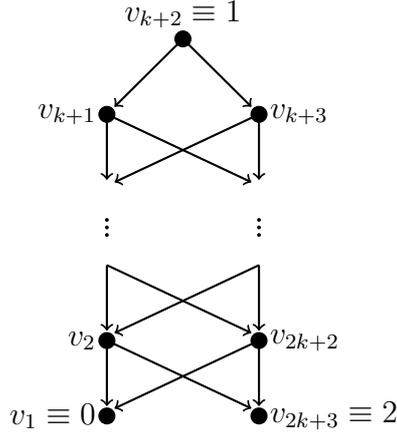
\end{cor}

Indeed, the above choice of $R(g)$ yields the equation
\begin{equation}
    v_t = v^2 \left[ v_{\theta\theta}+\frac{v_\theta}{\tan(\theta)}+ \lambda v(v-1)(2-v) \right]
\end{equation}
for $\lambda\in\mathbb{R}$. The unkown $w:=v-1$ satisfies the Chafee-Infante equation with quasilinear diffusion coefficient $(w+1)^2$. Hence, the equilibria $v\equiv 1$, corresponding to Schwazrschild, has the role of the bifurcating equilibria $w\equiv 0$ in the usual Chafee-Infante equation and each time $\lambda$ crosses an eigenvalue of the spherical Laplacian, the Schwarzschild solution bifurcates to an axisymmetric solution.

It would be interesting to compute the attractor for the prescribed scalar curvature from \cite{FiedlerHellSmith}, namely $R=(\lambda+2)/r^2$. This yields a slowly non-dissipative nonlinearity with non hyperbolic equilibria. That is, solutions $v$ might now stay bounded as $r\to r_1$ and grow-up occurs. This will be done in the near future.

Note that in order to construct the attractor $\mathcal{A}_{axi}$, one needs to know the \emph{zero number} of the difference of solutions $v-{v_{k+2}}$, where the trivial solution ${v_{k+2}}\equiv 1$ represents the Schwarzschild solution in self similar variables. Roughly speaking, one needs to know how many intersections other equilibria have with ${v_{k+2}}$. This encodes the information of how much such equilibria deviate from ${v_{k+2}}$, and whenever a solution intersects with the trivial solution, it means that $v(r,\theta)=1$ and the metric looks like the Schwarzschild solution at that fixed radius $r$.

The second main result regarding certain elements of the event horizon answers partially the question of how the symmetry of the sphere dictates a symmetry of some solutions in the attractor $\mathcal{A}$. This is done precisely in \cite{LappicySym}.

A function $v\in C^1(\mathbb{S}^2)$ has \emph{axial extrema} if its maxima and minima in $\phi$ occur as axis from the north to south pole. In other words, if $v_\phi(\theta_0,\phi_0)=0$ for a fixed $(\theta_0,\phi_0)\in \mathbb{S}^2$, then $v_\phi(\theta,\phi_0)=0$ for any $\theta\in[0,\pi]$. In that case, the extrema depend only at the position in $\phi$. Note that if $R(g)$ is analytic, then the solution $v$ of \eqref{INTRO:IVP} is also, as in \cite{CaoRammahaTiti00}. Then the set of axial extrema is finite, and we denote them by $\{ \phi_i \}_{i=0}^N$ where $\phi_0:=\phi_N$. 

Axial extrema are \emph{leveled} if all axial maxima $\phi_i$ have the same value $u(\theta,\phi_i)=M(\theta)$, and all axial minima $\phi_i$ also have the same value $u(\theta,\phi_i)=m(\theta)$.


\begin{cor}\emph{\textbf{Symmetry within the Event Horizon}} \label{I:SYMTHM}

Suppose the scalar curvature $R$ is analytic and $v(r_1,\theta,\phi)$ is an equilibrium of \eqref{INTRO:IVP} within the attractor $\mathcal{A}$ that only has axial extrema $\{ \phi_i \}_{i=0}^N$ where $\phi_0:=\phi_N$.

Then $\phi_{i}=(\phi_{i-1}+\phi_{i-1})/2$ and the self-similar Schwarzschild metrics at the event horizon have the following reflection symmetry
\begin{equation}\label{SYM2}
    v(r_1,\theta,\phi)=v(r_1,\theta,R_{\phi_i}(\phi))
\end{equation}
for all $i=1,...,N$, where $(\theta,\phi)\in[0,2\pi] \times [\phi_{i-1},\phi_{i}]$ and $R_{\phi_i}(\phi)=2\phi_i-\phi$.
\end{cor}

This theorem raises the mathematical question whether such result holds for other domains, such as the torus or the hyperbolic disk. Subsequently, it raises the physical question of space foliated by other two dimensional surfaces than the sphere, and if the resulting equation for the scalar curvature is still parabolic and of the same form as \eqref{EEINTRO}. Even though Hawking's theorem in \cite{Hawking1972} stating that event horizons for certain black holes are topologically $\mathbb{S}^2$, other stellar objects of interest could carry different topology. For example, it was found numerically that dust collapse might yield a toroidal horizon before reaching its spherical shape in \cite{Abrahamsetal94}. Therefore, it is expected that the set $\mathcal{X}$ from Corollary \ref{I:SPC} and the set of initial data for toroidal foliations are connected to each other in phase-space. 

Lastly, this theorem was achieved by trying to prove a symmetrization of Gidas, Ni and Nirenberg type: positive solutions of parabolic equations on the sphere are axial. This has been achieved for subsets of the sphere, see \cite{LappicySym} and references therein. If this conjecture of symmetrization on the sphere was true, we would obtain the the phase-space $X$ above, which consists of positive solutions, they constitute of axially symmetric functions. Therefore, $X=X_{axi}$ and $\mathcal{A}=\mathcal{A}_{axi}$.

\textbf{Acknowledgment.} The author is indebted to Bernold Fiedler for proposing this project for my Phd thesis on a walk towards the Elbe, in Wittenberg. This project was funded by the Berlin Mathematical School and CAPES [grant number 99999.009627].

\medskip


\end{document}